# Surface EMG Profiling in Parkinson's Disease: Advancing Severity Assessment with GCN-SVM

Daniel Cieślak[1, 3], Barbara Szyca[1], Michalina Razik[1], Liwia Florkiewicz[1], Anna Prus[2], Hubert Lis[2], Kinga Grzęda[1], Weronika Bajko[2], Helena Kamieniecka[2], Weronika Matwiejuk[2], Inga Rozumowicz[2], Wiktoria Ziembakowska[2] and Mariusz Kaczmarek[1]

[1] Gdańsk University of Technology, Gdańsk, Poland
[2] Medical University of Gdańsk, Gdańsk, Poland
IDEAS NCBR, Warsaw, Poland
milab.weti@student.pg.edu.pl

**Abstract.** Parkinson's disease (PD) poses challenges in diagnosis and monitoring due to its progressive nature and complex symptoms. This study introduces a novel approach utilizing surface electromyography (sEMG) to objectively assess PD severity, focusing on the biceps brachii muscle. Initial analysis of sEMG data from five PD patients and five healthy controls revealed significant neuromuscular differences. A traditional Support Vector Machine (SVM) model achieved up to 83% accuracy, while enhancements with a Graph Convolutional Network-Support Vector Machine (GCN-SVM) model increased accuracy to 92%. Despite the preliminary nature of these results, the study outlines a detailed experimental methodology for future research with larger cohorts to validate these findings and integrate the approach into clinical practice. The proposed approach holds promise for advancing PD severity assessment and improving patient care in Parkinson's disease management.

## 1 Introduction

Parkinson's disease (PD) is a progressive neurological disorder affecting a significant portion of the elderly population globally [4]. Marked by debilitating motor symptoms such as tremors, stiffness, bradykinesia, and postural instability, PD significantly impairs daily activities and reduces quality of life [9]. The disease's progressive nature and complex symptomatology present substantial challenges in its diagnosis and management. Although the Unified Parkinson's Disease Rating Scale (UPDRS) is employed to assess PD severity, its subjective and time-consuming nature in detecting early-stage symptoms are notable limitations [5]. In light of these challenges, wearable sensors, particularly surface electromyography (sEMG), offer promising solutions for objective PD assessment by providing detailed insights into neuromuscular function [6]. Previous studies have highlighted sEMG's effectiveness in evaluating motor symptoms and monitoring disease progression in PD patients, as reviewed in the *Related Work* section.

This study aims to develop and validate a novel sEMG-based methodology for objectively assessing PD severity, focusing on the biceps brachii muscle. By analyzing both linear and nonlinear sEMG parameters, we seek to identify neuromuscular differences between PD patients and healthy controls. Furthermore, we propose a Graph Convolutional Network-Support Vector Machine (GCN-SVM) model designed to achieve better and more reliable results on small data samples compared to classical methods.

This research contributes in two fundamental ways: firstly, by introducing a detailed and novel sEMG-based methodology for the objective evaluation of PD severity, emphasizing the analysis of both linear and nonlinear sEMG parameters. Secondly, through the introduction of the GCN-SVM model, which exhibits significant advancements in classification accuracy over traditional SVM approaches, particularly noteworthy given the constraints typically associated with pilot studies involving limited data. Despite the small sample size, our preliminary results indicate that the GCN-SVM model significantly enhances the accuracy and reliability of PD severity assessments. This innovative approach addresses the limitations of existing subjective tools and has the potential to improve clinical practices by providing a more individualized approach to therapy and enhancing PD management.

Subsequent sections provide detailed descriptions of the study cohort, experimental hardware configurations, and measurement protocols within the *Methods* section. Furthermore, this section discusses the data analysis techniques employed. The *Results* section presents comprehensive findings from the comparative analysis and performance evaluations of the proposed models. Finally, the *Conclusion* section provides a comprehensive summary of our findings along with the recommendations for future research to advance sEMG-based methodologies in Parkinson's disease assessment.

## 2   Related work

Advanced methods are crucial for accurate diagnosis and monitoring of Parkinson's disease. The integration of technology, particularly wearable sensors, has shown promise in enhancing clinical assessments and treatment strategies. This study builds upon previous research that has utilized sEMG to evaluate motor symptoms and disease progression in PD patients:

Fundaro et al. [1] compared sEMG recordings in healthy individuals and PD patients, finding significant correlations between sEMG findings and Parkinson Fatigue Scale (PFS) scores. They gathered demographic and clinical data, ensuring consistent L-DOPA dosage for at least two weeks before the experiment. The protocol included three types of movements.The study found significant correlations between sEMG findings and both PFS and UPDRS scores but no discernible difference in muscle strength between the groups.

Kleinholdermann et al. [5] developed a non-invasive method for assessing motor symptoms in 45 PD patients using sEMG from a wristband. Patients performed a tapping task at five-second intervals. The study found a correlation coefficient of 0.739 between sEMG activity and UPDRS scores, indicating the potential utility of sEMG features in predicting medication adjustments, particularly in telemedicine applications.

Meigal et al. [6] compared linear and nonlinear sEMG parameters across healthy young individuals, healthy elderly participants, and PD patients. The study revealed that RMS (Root Mean Square) increased with load in all groups, while MDF (Mean Frequency) showed a significant increase only in the PD group. Nonlinear parameters exhibited notable differences between the PD and control groups, especially without additional load. Significant correlations were established between novel sEMG parameters and UPDRS scores, highlighting the potential of these nonlinear parameters in assessing motor symptoms and disease severity in PD patients.

In recent years, Support Vector Machines (SVM) have been widely utilized in various medical applications, including Parkinson's disease assessment. SVMs have shown effectiveness in classifying and predicting disease severity based on different data modalities [7]. SVMs have been particularly valuable in handling small datasets and achieving high accuracy in disease prediction tasks [7]. However, the traditional SVM model may face limitations in capturing complex relationships within the data, especially in scenarios with limited sample sizes.

On the other hand, Graph Convolutional Networks (GCN) have emerged as a powerful tool for learning from graph-structured data, offering the ability to capture intricate relationships and dependencies within the data [8]. GCNs have been successfully applied in various domains, including brain disease diagnosis [8]. introduced a framework that employed GCN to explore deeper relationships for improved disease prediction [8]. GCNs have the advantage of aggregating information from neighboring nodes in a graph, enabling the model to capture local graph structures and identify patterns effectively [8]. In the context of PD assessment, GCNs can enhance the classification accuracy by leveraging the graph-based representations of sEMG data, which may contain complex interdependencies that traditional models like SVMs may struggle to capture.

The integration of GCN with SVM, as seen in the GCN-SVM model proposed in this study, represents a novel approach that combines the strengths of both methods. By leveraging the graph-based representation capabilities of GCNs and the classification power of SVMs, the GCN-SVM model aims to provide more accurate and reliable assessments of PD severity based on sEMG data [8].

## 3 Methods

In this section, we provide a comprehensive description of the study cohort, detailing the measurement methodology employed and outlining the approach used for data analysis. This study was approved by the Ethics Committee of Gdańsk University of Technology. All subjects gave their written informed consent before participating.

### 3.1 Study cohort

The study included five PD patients and five healthy controls. PD patients were recruited from the Association for People with Parkinson's Disease and Brain Degenerative Diseases and their Caretakers 'Park On' in Gdańsk, Poland. Inclusion criteria were prior PD diagnosis; exclusion criteria included other motor function-affecting diseases, pacemakers, metal implants, and recent upper limb injuries. The study group demographics and clinical characteristics are summarized in Table 1. Disease duration refers to the length of time since the individual was diagnosed with Parkinson's disease. Medication duration, in the context of Parkinson's disease, refers to the length of time the patient has been taking medications specifically prescribed to manage their Parkinson's symptoms. The control group was selected to match the age distribution of the PD group, ensuring comparability, as shown in Table 2.

**Table 1.** Characteristics of the study group.

| PD Patient | Age | Gender | BMI | Disease duration (years) | Medication duration (years) |
|---|---|---|---|---|---|
| 1 | 76 | F | 23,8 | 4 | 3 |
| 2 | 64 | M | 22,3 | 12 | 12 |
| 3 | 84 | M | 17 | 5 | 3 |
| 4 | 73 | F | 24,2 | 6 | 6 |
| 5 | 72 | F | 21,9 | 7 | 7 |

**Table 2.** Characteristics of the control group.

| Subject | Age | Gender | BMI |
|---|---|---|---|
| 1 | 67 | M | 26.60 |
| 2 | 71 | M | 31.46 |
| 3 | 71 | M | 24.91 |
| 4 | 64 | M | 29.38 |
| 5 | 80 | M | 33.00 |

### 3.2 Hardware

The measuring device employed in our study utilizes the Myoware 2.0 Muscle Sensor [3], designed for assessing the electric voltage associated with muscular contraction. This sensor interfaces with gel Ag/AgCl electrodes that comply with ANSI/AAMI EC12 standard for disposal ECG electrodes [2]. The electrodes were positioned bilaterally on the skin surface to capture electrical signals from the biceps brachii muscle, as illustrated in Fig. 1.

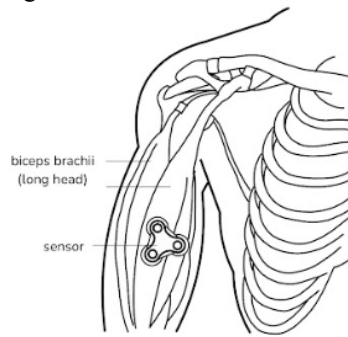

**Fig. 1.** Placement of the electrode on the biceps muscle.

The non-invasive methodological approach minimizes participant discomfort and preserves skin integrity. Signal acquisition is done via jumper wires connecting the sensor to an ESP32 board, which uses Bluetooth v4.2 for wireless data transmission. The system, powered by a 5V power bank, avoids signal interference issues common with PC power sources. The Myoware Sensor converts analog muscle activity signals into digital data, reflecting muscle contraction characteristics like amplitude, frequency, and waveform shape. Data is sent to a mobile device, then transferred to a computer for generating voltage-muscle time dependency graphs, offering insights into muscle activity patterns over time.

### 3.3 Measurement Procedure

To collect sEMG data, we devised a measurement procedure enabling the comparison of muscle activity between healthy individuals (control group) and subjects with Parkinson's disease (study group). The procedure includes three stages:

1. The first stage involves measuring the resting activity of the biceps muscle for 30 seconds.
2. The second stage involves a 30-second measurement where participants hold a 2 kg weight. The arm is positioned with the elbow bent at a 90-degree angle, close to the body, and the forearm parallel to the floor. This posture induces an isometric contraction, allowing assessment of muscle activity under load.
3. The final stage requires participants to perform alternating arm flexion and extension with the 2 kg weight at 5-second intervals. This dynamic movement pattern assesses muscle activity during functional tasks that mimic daily activities.

To ensure consistent timing and initiation of measurements, the servo mechanism was utilized. This mechanism, placed in participants' hands, generates vibrations to signal the start and end of each measurement stage. Additionally, messages printed with timestamps during each stage facilitate data processing synchronization. Before each stage begins, participants receive three servo vibrations as a readiness confirmation. The measurement itself starts and stops with a single vibration, and each interval within the stage is also marked by a vibration signal.

The servo mechanism was integral to our protocol, ensuring standardized timing cues and minimizing variability in data collection. By providing clear signals to participants and autonomously managing measurement initiation, the servo mechanism enhanced the reliability and accuracy of our sEMG data, crucial for comparisons between healthy individuals and those with Parkinson's disease.

### 3.4 Data processing

Python was used for EMG data analysis, leveraging libraries such as Pandas, NumPy, SciPy, Matplotlib, and Seaborn. The analysis process included data collection, cleaning, feature extraction, visualization, and model evaluation to ensure data integrity and quality.

*Data Collection and Cleaning*
The initial phase of the data processing pipeline involved the collection and cleaning of acquired sEMG data, stored in CSV files. These files contained recordings from both healthy individuals and subjects diagnosed with Parkinson's disease, including timestamps and muscle activity values measured in Arbitrary Units (AU). To ensure the integrity and quality of the data, validation checks were performed during the collection process. This involved confirming the absence of outliers and ensuring continuous sampling intervals. Any measurements that fell outside the expected range were either corrected or the measurement was repeated, ensuring the completeness and accuracy of the dataset.

Subsequently, the collected data were systematically organized into distinct folders based on the health condition of the subjects (PD or healthy). The example of an acquired sEMG signal is shown in Fig. 2. The signal has five marked segments that represent different phases of muscle activity: Segment 1 - Relaxation Phase; Segment 2 - Preparation for Contraction; Segment 3 - Maintenance of Contraction; Segment 4 - Preparation for Repetitions; Segment 5 - Execution of Repetitions.

In the figure, the x-axis represents time in seconds (s), while the y-axis denotes the result in Arbitrary Units (AU), reflecting the muscle activity level recorded by the sEMG sensor. Segment 1 shows the muscle in a relaxed state with minimal activity. Segment 2 indicates a slight increase in muscle activity as the participant prepares for contraction. Segment 3 captures sustained muscle activity while holding a 2 kg weight, indicated by increased and stable sEMG signals. Segment 4 involves a brief preparatory period before performing repetitive movements, with a slight drop in muscle activity. Segment 5 shows periodic spikes in the sEMG signal as the participants perform alternating arm flexion and extension.

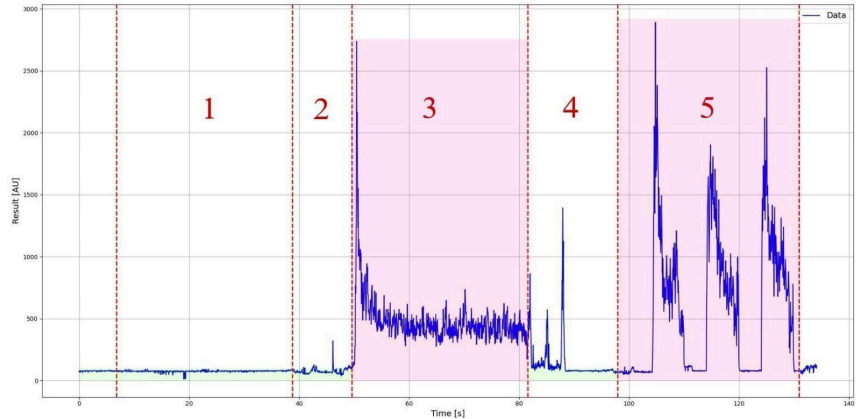

**Fig. 2.** Example of an sEMG signal with marked segments.

Further categorization was done by hand dominance (left or right). Specific segments of data corresponding to the exercise phases (1, 3, and 5) were isolated for detailed analysis. This structured approach facilitated a streamlined and efficient data processing workflow.

*Feature Extraction*

Various statistical features were extracted from the sEMG signals to facilitate the analysis. Key features included Root Mean Square (RMS), Median Frequency (MDF), Mean Frequency, Variance, Skewness, Kurtosis, Standard Deviation (SD), Percentage of Recurrence (REC), Percentage of Determinism (DET), Sample Entropy (SampEn), and Correlation Dimension (CD). These features were chosen based on their established relevance in the literature for characterizing neuromuscular activity.

RMS was calculated to provide a measure of the signal's amplitude, while MDF and Mean Frequency offered insights into the frequency characteristics of the sEMG signals. Variance, Skewness, and Kurtosis described the statistical distribution of the signal values, highlighting their spread, asymmetry, and peakedness, respectively. SD was used to quantify the dispersion of the signal data points. REC and DET, derived from recurrence quantification analysis, provided information on the repeating patterns within the sEMG signals, which is critical for understanding muscle activity regularity. SampEn quantified the complexity and irregularity of the signals, with higher values indicating more complex muscle activation patterns. CD measured the signal's complexity in terms of fractal dimensions.

The extraction process ensured that these features were consistently calculated across all samples, facilitating a reliable comparison between healthy individuals and PD patients. This comprehensive feature set enabled a detailed analysis of the neuromuscular differences between the two groups, providing a robust basis for subsequent model training and evaluation.

### 3.5    **GCN-SVM Architecture and Implementation**

The GCN-SVM model effectively combines GCN and SVM to classify Parkinson's disease using sEMG data. The process involves detailed data preprocessing, feature extraction, and graph construction, as shown in Table. 3. The GCN model, with its layered architecture, captures complex relationships in the data, leading to improved classification performance. Training and evaluation involve cross-entropy loss, Adam optimization, and extensive cross-validation to credibility.

**Table 3.** Detailed GCN-SVM Model Architecture.

| Stage | Description | Details |
|---|---|---|
| Feature Calculation | Extract statistical features from sEMG signals. | RMS, MDF, Mean Frequency, Variance, Skewness, Kurtosis, Max, Min, Range. Fourier Transform (FFT) used for frequency domain features. |
| Data Loading | Load and clean sEMG data from CSV files. | Label data for 'healthy' and 'PD' groups. Separate by hand (left/right) and patient segments. Standardize features using StandardScaler. |
| Graph Construction | Create a KNN graph to represent relationships between data points. | Use kneighbors_graph to construct the graph. Convert adjacency matrix to edge indices. |
| GCN Model Architecture | Graph Convolutional Network layers for feature processing. | GCNConv Layer 1: Initial graph convolution. ReLU Activation: Introduces non-linearity. Dropout Layer: Regularization. GCNConv Layer 2: Second graph convolution. ReLU Activation: Non-linearity. Dropout Layer: Regularization. GCNConv Layer 3: Final graph convolution. Log Softmax: Output classification. |
| Model Initialization | Initialize model, optimizer, and loss function. | Model initialized with 16 hidden channels and 2 output channels. Adam optimizer with learning rate of 0.01. Cross-entropy loss (NLLLoss). |
| Training | Train the model with the training dataset. | Perform forward and backward passes. Update model weights using optimizer. Train for 50 epochs. |
| Cross-Validation | Evaluate model performance using cross-validation. | Stratified K-Fold cross-validation to assess model performance on different subsets. |
| Testing | Evaluate the trained model on a test dataset. | Assess model accuracy, generate confusion matrices, and create classification reports. |
| Data Dimensions | Various dimensions of data and model inputs. | Feature Dimensions: 9 (RMS, MDF, MeanFreq, Variance, Skewness, Kurtosis, Max, Min, Range). Hidden Channels: 16. Output Channels: 2 (Healthy/PD). |

| Performance Metrics | Metrics to evaluate model performance. | Accuracy, F1-Score, Precision, Confusion Matrix. |

## 4 Results

The results section presents a detailed analysis of the collected sEMG data, comparing the neuromuscular characteristics between Parkinson's disease patients and healthy controls. Additionally, it evaluates the performance of the proposed Graph Convolutional Network-Support Vector Machine (GCN-SVM) model in accurately classifying PD severity based on the extracted features.

### 4.1 Linear and Non-Linear sEMG parameters

The analysis identified several key sEMG parameters, both linear and non-linear. Among the linear parameters, Median Frequency (MDF) showed stability across samples, making it a critical metric for distinguishing between healthy and PD-affected muscles. MDF demonstrated variability in frequency domain characteristics, reflecting notable differences between groups. Mean Frequency values hovered around zero, indicating a balanced frequency distribution essential for reliable comparative analysis. Root Mean Square (RMS) had a mean of 503.08 (SD = 491.26, range = 12.84 to 2548.65), indicating a wide range of values.

Non-linear parameters also showed significant variability. Skewness had a mean of 2.47 (SD = 3.53, range = -4.52 to 19.85), while Kurtosis had a mean of 24.83 (SD = 65.38, range = -1.57 to 426.37). Sample Entropy (SampEn) highlighted the complexity and irregularity of EMG signals, with higher values indicating more complex muscle activation patterns typical in PD. Our study corroborates the findings of Meigal et al. [6], showing similar trends in RMS and MDF with higher values in the PD group. Table 4 summarizes the statistical characteristics of the features calculated from the EMG signals.

**Table 4.** Statistical characteristics (PD vs Healthy)

| Feature | Mean | | Std | | Min | | Max | |
|---|---|---|---|---|---|---|---|---|
| | PD vs Healthy | | PD vs Healthy | | PD vs Healthy | | PD vs Healthy | |
| RMS | 503.08 | 466.99 | 491.26 | 322.04 | 12.84 | 27.64 | 2548.65 | 961.53 |
| MDF | 7.69 | 6.87 | 1.91 | 3.27 | 0.31 | 0.31 | 9.75 | 9.75 |
| MeanFreq | -0.004 | -0.002 | 0.006 | 0.006 | -0.24 | -0.014 | 0.00 | 0.004 |
| Skewness | 2.47 | 2.69 | 3.53 | 4.05 | -4.52 | -4.52 | 19.85 | 10.17 |
| Kurtosis | 24.83 | 11.59 | 65.38 | 43.89 | -1.57 | -0.84 | 426.37 | 117.03 |
| SD | 244.71 | 239.39 | 237.57 | 199.52 | 2.02 | 2.02 | 1192.09 | 658.09 |
| REC | 0.032 | 0.048 | 0.048 | 0.065 | 0.003 | 0.003 | 0.2 | 0.2 |
| DET | 0.255 | 0.077 | 1.285 | 0.089 | 0.00 | 0.00 | 10.239 | 0.27 |
| SampEn | 0.898 | 0.684 | 0.614 | 0.512 | 0.044 | 0.044 | 2.306 | 1.76 |
| CD | 1.29 | 1.06 | 0.56 | 0.64 | 0.00 | 0.00 | 2.09 | 2.107 |

The presented results contribute to understanding the diverse characteristics of sEMG signals in both healthy individuals and those with Parkinson's disease, emphasizing the importance of considering a broad range of muscle activities and conditions in such studies [6]. The classification model's effectiveness in distinguishing between healthy individuals and those with Parkinson's disease was evaluated across different data subsets (Table 5).

Table 5. Classification model performance for the complete dataset, right hand data, and left hand data.

| Metric | Complete dataset | Right hand data | Left hand data |
|---|---|---|---|
| Cross-Validation Scores | [0.772, 0.777, 0.766, 0.774, 0.773] | [0.784, 0.846, 0.825, 0.835, 0.828] | [0.811, 0.806, 0.802, 0.804, 0.804] |
| Confusion Matrix | [[3519, 830], [1005, 2877]] | [[1863, 282], [410, 1553]] | [[1755, 422], [379, 1567]] |
| F1-Score (Healthy) | 0.79 | 0.84 | 0.81 |
| F1-Score (PD) | 0.76 | 0.82 | 0.80 |
| Accuracy | 0.78 | 0.83 | 0.81 |
| Macro Avg F1-Score | 0.78 | 0.83 | 0.81 |
| Weighted Avg Precision | 0.78 | 0.83 | 0.81 |

The discrepancies in the number of samples between the right hand, left hand, and the complete dataset arise from the nature of the data collection process. Each participant had multiple measurements taken, resulting in several samples per user. The numbers reflect the total measurements taken for each hand separately. In contrast, the complete dataset combines all measurements from both arms.

These results indicate that the classification model performs better with the right-hand data compared to the left-hand data and the complete dataset, as evidenced by higher cross-validation scores and F1-scores.

### 4.2 GCN-SVM Classification Effectiveness of sEMG in PD Assessment

The effectiveness of the GCN-SVM classification model in distinguishing between healthy individuals and those with Parkinson's disease was evaluated across different subsets of the data (Table 6). Despite being based on a small sample size, these preliminary results show promising accuracy and reliability, significantly better than the classic SVM. Graph statistics provide insights into the structural properties of the graphs used in the GCN model. Key statistics include an average degree of approximately 9.964 across all folds, indicating robust connectivity; an average centrality value consistently at 0.000303, suggesting uniform centrality across nodes; and a high average clustering coefficient of 0.991021, reflecting strong clustering and implying that nodes form tight-knit groups.

Table 6. GCN-SVM model performance for the complete dataset

| Metric | Healthy vs PD subjects |
|---|---|
| Cross-Validation Scores | [0.843, 0.931, 0.937, 0.949, 0.946] |
| Confusion Matrix | [[4003. 346], [302. 3578]] |
| F1-Score (Healthy) | 0.93 |
| F1-Score (PD) | 0.91 |
| Accuracy | 0.92 |
| Macro Avg F1-Score | 0.92 |
| Weighted Avg Precision | 0.93 |

The GCN-SVM model significantly outperforms the traditional SVM model in classifying healthy individuals and those with Parkinson's disease. The GCN-SVM achieved a higher accuracy (**0.92 vs. 0.78**), improved F1-scores (**0.93 vs. 0.79** for healthy, **0.91 vs. 0.76** for PD), and better precision. The enhanced performance of GCN-SVM can be attributed to its ability to capture complex relationships and dependencies within the sEMG data through graph-based representations.

In the "Cross-Validation Scores" row, accuracy is used as the primary metric, defined as the ratio of correct predictions to the total number of test examples, expressed as a percentage or fraction. The process involves 5-fold cross-validation with StratifiedKFold, ensuring class proportions are maintained across folds. The data is divided into five equal parts, and in each iteration, the model is trained on four parts and tested on one part. This is repeated five times with different test folds. Individual fold results are recorded, and the final cross-validation result is the average accuracy across all folds. This simplicity and intuitiveness make accuracy an ideal metric for reporting.

## 5  Conclusion and Future works

This study introduced a novel approach utilizing surface electromyography for objectively assessing Parkinson's disease severity, focusing on the biceps brachii muscle. By analyzing both linear and nonlinear sEMG parameters, significant neuromuscular differences between PD patients and healthy controls were identified. The findings underscored the utility of sEMG as a non-invasive tool for precise PD severity assessment, particularly when combined with advanced machine learning techniques.

Key linear parameters such as Median Frequency exhibited stability across samples. Non-linear parameters (Skewness and Kurtosis) demonstrated variability, highlighting their role in capturing the complexity of muscle activation patterns.

Moreover, the study proposed and evaluated a GCN-SVM model, which outperformed traditional SVM approaches, particularly in scenarios with limited data samples. The GCN-SVM model achieved higher accuracy, improved F1-scores, and superior precision by effectively modeling the intricate relationships within sEMG data through graph-based representations. Future research directions should prioritize validating these findings with larger, diverse cohorts to enhance the robustness and generalizability of the proposed methodology. Such efforts will further refine sEMG-based assessments, potentially paving the way for personalized therapeutic strategies and improving overall patient care in PD management.

In conclusion, the integration of sEMG with advanced machine learning methodologies represents a significant advancement in Parkinson's disease research, offering nuanced insights into motor dysfunction and laying the foundation for tailored therapeutic interventions aimed at optimizing patient outcomes.

**Acknowledgments.**
This study was made possible by the invaluable support of The Association for People with Parkinson's Disease, Degenerative Brain Disorders, and Caregivers - 'Park On' in Gdańsk. Additionally, the research was supported by the Dean of the Faculty of ETI, benefiting the Medical Intelligence Laboratory.

# References

1. Fundaró, C., Gazzoni, M., Pinna, G., Dallocchio, C., Rainoldi, A., & Casale, R. (2021). Is fatigue a muscular phenomenon in Parkinson's disease? Implications for rehabilitation. European Journal of Physical and Rehabilitation Medicine, 57. https://doi.org/10.23736/S1973-9087.21.06621-1
2. Fannin. (2024). ECG Catalogue. Retrieved April 30, 2024, from https://www.fannin.eu/wp-content/uploads/2018/06/ECG-Catalogue.pdf
3. MyoWare. (2024). Technical Specifications: Muscle Sensor. Retrieved April 22, 2024, from https://myoware.com/products/technical-specifications/#muscle-sensor
4. World Health Organization. (2024). Parkinson disease. Retrieved April 22, 2024, from https://www.who.int/news-room/fact-sheets/detail/parkinson-disease
5. Kleinholdermann, U., Wullstein, M., & Pedrosa, D. (2021). Prediction of motor Unified Parkinson's Disease Rating Scale scores in patients with Parkinson's disease using surface electromyography. Clinical Neurophysiology, 132(7), 1708–1713. https://doi.org/10.1016/j.clinph.2021.01.031
6. Meigal, A. I., Rissanen, S., Tarvainen, M. P., Karjalainen, P. A., & Airaksinen, O. (2009). Novel parameters of surface EMG in patients with Parkinson's disease and healthy young and old controls. Journal of Electromyography and Kinesiology, 19(3), e206–e213. https://doi.org/10.1016/j.jelekin.2008.02.008
7. Mudawi, N. (2024). Developing a model for Parkinson's disease detection using machine learning algorithms. Computers Materials & Continua, 79(3), 4945-4962. https://doi.org/10.32604/cmc.2024.048967
8. Zhou, H., & Zhang, D. (2021). Graph-in-graph convolutional networks for brain disease diagnosis. Proceedings of the IEEE International Conference on Image Processing. https://doi.org/10.1109/icip42928.2021.9506259
9. Martignon, C., Laginestra, F. G., Giuriato, G., Pedrinolla, A., Barbi, C., Vico, I. A. D., & Venturelli, M. (2021). Evidence that neuromuscular fatigue is not a dogma in patients with Parkinson's disease. Medicine & Science in Sports & Exercise, 54(2), 247-257. https://doi.org/10.1249/mss.0000000000002791